# Non-linear vortex dynamics and transient effects in ferromagnetic disks

X.M. Cheng[1], K. Buchanan[2], R. Divan[3], K.Y. Guslienko[4], D.J. Keavney[1]

[1] *Advanced Photon Source, Argonne National Laboratory, Argonne, IL USA*

[2] *Department of Physics, Colorado State University, Fort Collins, CO USA*

[3] *Center for Nanoscale Materials, Argonne National Laboratory, Argonne, IL USA*

[4] *Dept. Materials Physics, University of the Basque Country 20080, San Sebastián, Spain*

**Abstract**

We report a time resolved imaging and micromagnetic simulation study of the relaxation dynamics of a magnetic vortex in the non-linear regime. We use time-resolved photoemission electron microscopy and micromagnetic calculations to examine the emergence of non-linear vortex dynamics in patterned $Ni_{80}Fe_{20}$ disks in the limit of long field pulses. We show for core shifts beyond ~20-25% of the disk radius, the initial motion is characterized by distortions of the vortex, a transient cross-tie wall state, and instabilities in the core polarization that influence the core trajectories.

In patterned thin film ferromagnets, demagnetizing fields favor an in-plane magnetization direction and a configuration that closes the magnetic flux within the element. In disk-shaped structures this can form a magnetic vortex state in which the local magnetization circulates in-plane, coupled with an out-of-plane core resulting from short range exchange interactions [1,2,3]. Recent interest in magnetoelectronic effects has fueled work in creating new high-speed spin-electronic technologies using patterned structures. Thus, the time-dependent magnetization processes exhibited by these objects are at the forefront of magnetism research [4]. Consequently, the influence of shape and finite size on magnetization dynamics in the Gigahertz frequency range has taken on a new importance.

Magnetic vortices exhibit a rich spectrum of magnetic excitations [5,6,7]. The fundamental mode corresponding to the translational, or gyrotropic, motion of the vortex core is of special interest [8,9]. For a circular structure with a moderate in-plane field applied, more spins will align along the field, resulting in a shift of the vortex core position perpendicular to the field. When the field is removed, the core relaxes towards the disk center [10,11]. Analytical and numerical calculations based on the Landau-Lifshitz-Gilbert (LLG) equation of motion predict that the core should spiral towards the disk center with an oscillation frequency characteristic of the lowest frequency eigenmode [6], with the sense of rotation being determined by the vortex core polarity.

The gyrotropic motion of the vortex core has been studied in several experiments using time-resolved x-ray photoemission microscopy (X-PEEM). Circular core motion was observed in Co squares under resonant excitation [12], while linear motion perpendicular to the applied field was observed in permalloy structures for excitation below [11] and above [13] the expected lowest eigenfrequency. Since these experiments involved samples of disparate size, shape, composition, and applied field amplitude, questions exist regarding the influences of these factors on the core trajectory. While some theoretical work suggests that elliptical trajectories are promoted by excitation far from resonance [14], changes to the internal structure of the vortex core during its motion may also influence the core trajectory and introduce a dependence on the field amplitude. Micromagnetic calculations suggest that the moving core becomes distorted to a degree dependent on its displacement from the disk center. These distortions may include regions of reversed core polarity and the possibility of core reversal via in-plane field excitation [15,16,17,18].

In this Letter, we describe a time-resolved X-PEEM and micromagnetic simulation study of the vortex core structure and trajectories in 6-$\mu$m-diameter permalloy disks, in which we image the undriven core motion after the removal of an *effectively steady-state* in-plane field. Our present investigations explore the vortex dynamics beyond the linear regime. We pay particular attention to the magnetic configuration within the first 1 ns after a field step, and its effects on the subsequent vortex core trajectories and dynamics. We observe a critical oscillation radius that corresponds to an *apparent* transition of the long-time core trajectory from circular to elliptical as the field is increased. This transition is also linked to the onset of magnetic vortex distortions associated with the vortex core polarity reversal.

Time-resolved PEEM experiments were performed at the Advanced Photon Source (APS) using samples made at the Center for Nanoscale Materials; both are located at Argonne National Laboratory. Coplanar waveguides were first fabricated on high resistivity Si wafers by photolithography, sputter deposition, and lift-off techniques [11]. Six-micron-diameter (2R), 30 nm-thick $Ni_{80}Fe_{20}$ disks were fabricated on top of the central conductor of the waveguide via electron beam lithography, sputter deposition, and lift-off processes. This geometry promotes the formation of vortices as the ground state, with a fundamental mode frequency that allows study of the non-resonant dynamics within the timing structure of the APS. Time-resolved PEEM magnetic images were taken at the Ni $L_3$ edge (852.7 eV) at APS beamline 4-ID-C in a pump-probe arrangement [11]. The pump pulses were 76ns long to provide a stable initial state, and separated by ~77ns zero-field intervals for study of the free vortex motion. The typical fall time was 150ps. The experimental data were compared to micromagnetic simulations [19], where the circular dots were represented using material parameters appropriate for bulk permalloy: saturation magnetization, $M_s$ = 8.0 x $10^5$ A/m, exchange stiffness constant $A$ = 1.3 x $10^{-11}$ J/m, and the magnetocrystalline anisotropy was neglected. The equilibrium states were obtained for each perturbation field using a large damping parameter ($\alpha$ = 1) for fast convergence, and then the magnetization state in each case was allowed to relax on removing the field using a more realistic damping parameter of $\alpha = 0.01$, along with a gyromagnetic ratio of $\gamma = 1.76 \times 10^2$ GHz/T. Simulations of a 6 μm-diameter disk were carried out for fields of 1, 2 and 4 mT using cells of 6x6 nm$^2$. To provide a more detailed look at the core behavior while maintaining reasonable computation times, simulations of a 3 μm-diameter disk were performed with 3x3 nm$^2$ cells (see Supplementary Material).

To examine the free oscillations of the vortex core with a pump-probe arrangement, we follow the core motion in the zero-field interval after the pump pulses. In Fig. 1, we show the core positions parallel and perpendicular to the applied field as a function of time after the falling step edge for 1, 2 and 4 mT fields. For the 1 mT excitation field, the initial displacement is ~0.15R, followed by a damped circular motion of the core towards the disk center, as indicated by the X and Y oscillations with a 90° phase difference. At 2 mT the initial displacement is ~0.4R, and the subsequent motion is elliptical (see Video 1 in Supplementary Material). When the bias field is increased to 4 mT, the initial displacement is ~0.7R, and the ellipticity is increased, with only small oscillations observed in the X-direction (see Video 2 in Supplementary Material). Both components oscillate at a frequency of 42.2 MHz, regardless of excitation field, which agrees well with the frequency of 42 MHz predicted by the LLG equation [6]. In the inset, we show the ratio of the X and Y amplitudes vs. field amplitude, which indicates that the X oscillations are small above 2.5mT. The observation of oscillations primarily in the Y direction after large core deflections is inconsistent with the expected circular vortex gyromotion [6], in which the motions along X and Y should be equal. After the initial displacement, the amplitude of the free Y oscillations is ~0.2R, for all bias field magnitudes ≥2 mT.

Thus at moderate bias fields, the core trajectory follows the circular path predicted in Ref. 6 and observed in Ref. 12, while for higher fields we observe a highly elliptical path with major axis perpendicular to the applied field, reminiscent of experiments done at higher fields [11,13,20]. A more detailed study of the spin configuration within the disk during the first 1 ns after the falling edge provides clues regarding the origin of the observed field dependence. In Fig. 2, we show several images taken just before and shortly after the falling edge of the 4 mT field pulse. Before the edge (Fig. 2a), a large displacement of the vortex core is evident, such that the core is almost expelled from the disk. Within 300 ps of the step edge, a domain state emerges that is not consistent with a simple vortex structure (Fig. 2b-d), characterized by a central region of gray contrast and a narrowing of the dark region of the disk. Experimentally, this domain state lasts only ~300 ps (Fig. 2e-j). Figure 3 shows the results of micromagnetic simulations of the core positions and spin distributions. We find a very similar Y-shift time dependence (Fig. 3a), as well as a very similar magnetic state (Fig. 3b) just after the falling edge of the pulse. Regions that were originally aligned along the field direction rapidly tilt such that the X-oriented portion of the vortex becomes smaller and the Y-oriented portions expand rapidly (Fig 2k). The position of the original vortex core actually changes little during the first ~2 ns. Instead, it becomes the top end of a complex cross-tie domain wall that extends towards the dot center. Thus, we postulate that the gray regions in the PEEM images result from the rapid expansion of small regions of Y-oriented spins into two large domains, one angled up and one down with a domain wall extending from the former core location to the top of the dark triangular domain. In the simulations, the domain wall consists of X-oriented spins, intermixed spins pointing out-of-plane both parallel and antiparallel to the original core polarity. The region oriented antiparallel to the field pulse expands as the spins relax back into a vortex configuration centered on the lower extent of the domain wall. During this process, multiple vortex-antivortex pair formations and polarity reversals occur along the wall. This can be seen in Fig. 3b, and in Video 3 in the Supplementary Material, which shows the simulated wall evolution for a 3 μm disk. Since a fresh vortex core forms at the new location, the polarity of the vortex is effectively uncorrelated with the original polarity. The transient cross-tie wall is similar to the predictions of micromagnetic simulations for large core deflections as the core is expelled from a micron-sized structure [21].

The existence of this transient domain state in the initial motion from highly displaced vortices provides valuable insight into the connection between non-linear vortex dynamics and core polarity reversals. The amplitude of the free oscillations in the experiments does not exceed ~0.2R, which is also the radius beyond which the transient states and elliptical trajectories appear. This suggests 20% of the disk radius as a limit to a linear regime in which the vortex core equation of motion is valid. In the simulations, the amplitude of the core circulation is ~0.15$M_s$, which corresponds to ~0.25$R$, in agreement with the experiment. If the core is displaced beyond this limit, distortions of the core region occur, promoting the transient domain state and subsequent randomization of the core polarity.

The vortex trajectories are well-described in the linear regime by analytical theory based on the Thiele's equation of motion [10] for the vortex core position $\mathbf{X}(t)$ [6,11]. The critical core trajectory radius and critical field pulse magnitude $H_{cr}$ can be estimated using the vortex $p$-reversal critical velocity $\upsilon_c$ [22], and the vortex annihilation field, $H_{an}$. The polarity reversal occurs when the vortex velocity $\upsilon \approx \omega_n |\mathbf{X}|$ is equal to the critical velocity $\upsilon_c \approx \gamma\sqrt{2A}$ (≈300 m/s for our parameters). Here $\omega_n(|\mathbf{X}|)$ is the non-linear vortex translation frequency accounting for quadric terms in the vortex energy, which is higher than the linear frequency $\omega$. Adding the nonlinear term $(\beta/4)\mathbf{X}^4$ to the vortex energy $W(\mathbf{X}) = \kappa \mathbf{X}^2/2$ [6,11] above, we get the nonlinear vortex frequency $\omega_n(|\mathbf{X}|) = \omega\left[1 + (\beta/\kappa)\mathbf{X}^2/R^2\right]$, $\beta$>0, where $\kappa$ is an effective stiffness coefficient. The ratio $\beta/\kappa$ is ≈ 4 within the pole-free model of the shifted vortex [6,8]. After the end of the field pulse of amplitude $H$, the core shift is $|\mathbf{X}|/R \approx \chi H/M_s$, where $\chi$ is the static susceptibility [6]. The value of $\chi^{-1}M_s = H_{an}$ can be found within the "rigid" vortex model [6] for 2$R$=6 $\mu$m, thickness $L$=30 nm, $M_s$=8.0 x 10$^5$ A/m to be ≈ 9.6 mT. Using the experimental value $\omega/2\pi$ = 42.2 MHz we find that the critical core trajectory radius is $|\mathbf{X}|/R \approx 0.28$, which agrees well with the simulations (0.25) and experiment (0.2). Solving the equation $\omega_n(|\mathbf{X}|)|\mathbf{X}| = \upsilon_c$, we get $H_{cr}/H_{an} \approx 0.28$ ($H_{cr}$≈ 2.7 mT). The measured value of $H_{cr}$ ~ 2.5 mT, taken as the field above which the free oscillation Y-amplitude no longer increases as a function of field pulse amplitude, is in reasonably good agreement with the estimated $H_{cr}$. Although the critical core shift is well-defined, the transition from circular to elongated core trajectories is gradual, as shown in the Fig. 1 inset. Our experimental critical core displacement value (0.2$R$) corresponds to an average vortex core velocity of

≈160 m/s. Similar to the measurements of Curcic et al. [23], this is lower than the estimate obtained from theory [22]. However we note that this value is the velocity after all transient motion has been damped out, and that during the initial core motion the velocity can be much higher.

The randomization of the core polarity suggests a mechanism to reconcile the observations of gyrotropic and non-gyrotropic trajectories seen here and in the literature [11,12,13,24]. The reformation of the vortex core with random polarity has the consequence that in the following gyrotropic motion the X-component of the vortex motion will cancel, giving the illusion of absence of motion in this direction, while the phase of the Y-component is maintained between pulses. This is suggested in our experiments by the broadening in the X direction of the core at $t = 8.1$ ns, as expected for an image resulting from the superposition of left and right shifted cores. Additional experiments show that the core circulation direction observed at 1 mT can be randomized by momentary pulsing at 4 mT. The micromagnetic simulations confirm that the motion following a single pulse is circular regardless of the magnitude of the field pulse (Fig. 3a). The simulations also show that above the critical field, the rising and falling edges of each pulse trigger at least one core reversal event as in Ref. [15]. For larger field pulses a new core forms far away from the old one such that the final core polarity is uncorrelated with the first, where the intermediate state is reminiscent of the complex cross-tie walls observed by Neudert et al. [25].

Thus we conclude that the free motion of a well-formed magnetic vortex always corresponds to a circular core trajectory, but that a threshold of ~0.2R exists beyond which transient distortions of the vortex core lead to non-linear dynamical effects. These distortions can give the appearance of a linear or strongly elongated core trajectory in pump-probe experiments through pulse-to-pulse randomization of the phase in the X motion. The repeatability of these measurements implies that the probability of core reversal from pulse to pulse is high, in agreement with the micromagnetic simulations. Finally, we note that the appearance of an elliptical trajectory requires that the probability of core reversal have values other than exactly 0 or 1. Such a core polarity bias may be induced by sample shape irregularities.

Use of the Advanced Photon Source and the Center for Nanoscale Materials at Argonne National Laboratory is supported by the U. S. Department of Energy, Office of Science, Office of Basic Energy

Sciences, under Contract No. DE-AC02-06CH11357. K.G. acknowledges support by the Ikerbasque Science Foundation.

Figure Captions:

Figure 1 (color online). Vortex core positions parallel (X) and perpendicular (Y) to the applied field vs. time, along with fits (red lines) to damped sinusoids after 1 (a), 2 (b) and 4 mT (c) field pulses. (d) shows the time-dependent applied field profile: 75-ns long field pulses, with falling edge at t=0. The upper inset shows the experimental geometry indicating the relative field (**B**) and photon momentum (**k**) directions. The lower inset shows the X deflection/Y deflection amplitude ratio vs. the pulse field amplitude.

Figure 2 (color online). (a-j) PEEM images at selected times after an applied field of 4 mT is removed. Between 0.3 and 0.6 ns (b-d), a domain state is observed as indicated by the gray regions in the left and right lobes of the disk. Between 0.9 and 1.9 ns (e-g), this domain state is not seen, possibly due to stochasticity in the domain configuration. After ~2 ns, a new vortex core forms at ~0.2R above the disk center and gyrotropic motion begins. At 8.1 ns, when the core should have its maximum shift in X, the image instead shows a core that is centered but broadened in the X direction. The inset shows a detail of the transient domain structure (k) and for an unshifted vortex (l).

Figure 3 (color online). Micromagnetic simulations of the vortex dynamics. (a) Plots of $M_x$ (solid) and $M_y$ (dotted) components as a function of time show counterclockwise circular vortex motion after a 1 mT pulse and clockwise vortex motion after 2 and 4 mT pulses. Each simulation was initiated with a vortex of positive polarity. The magnetic field pulse shape is shown as a black dash-dot line. (b) Images of the magnetization distribution before and after the 4 mT field pulse are shown, where red (blue) represent the magnetization component aligned (anti-aligned) with x. Cross-sectional plots of $M_z$ along y reveal that the vortex core is replaced by a complex cross-tie domain wall within 0.6 ns of the pulse, consisting of spins of both positive and negative polarity. At 8 ns, $M_z$ vs. y shows the single, new core of negative polarity.

[24] In Ref. 13, the frequency of excitation was 62.5 MHz. Additional data taken on permalloy squares of the same size and thickness using our low-frequency excitation scheme show that the gyrotropic mode frequency is ~35 MHz. Thus, the core trajectory should be elongated rather than circular (see Lee and Kim, Phys. Rev. B 78, 014405 (2008)).

[25] A. Neudert, et al., *Phys. Rev. B* **75**, 172404 (2007).

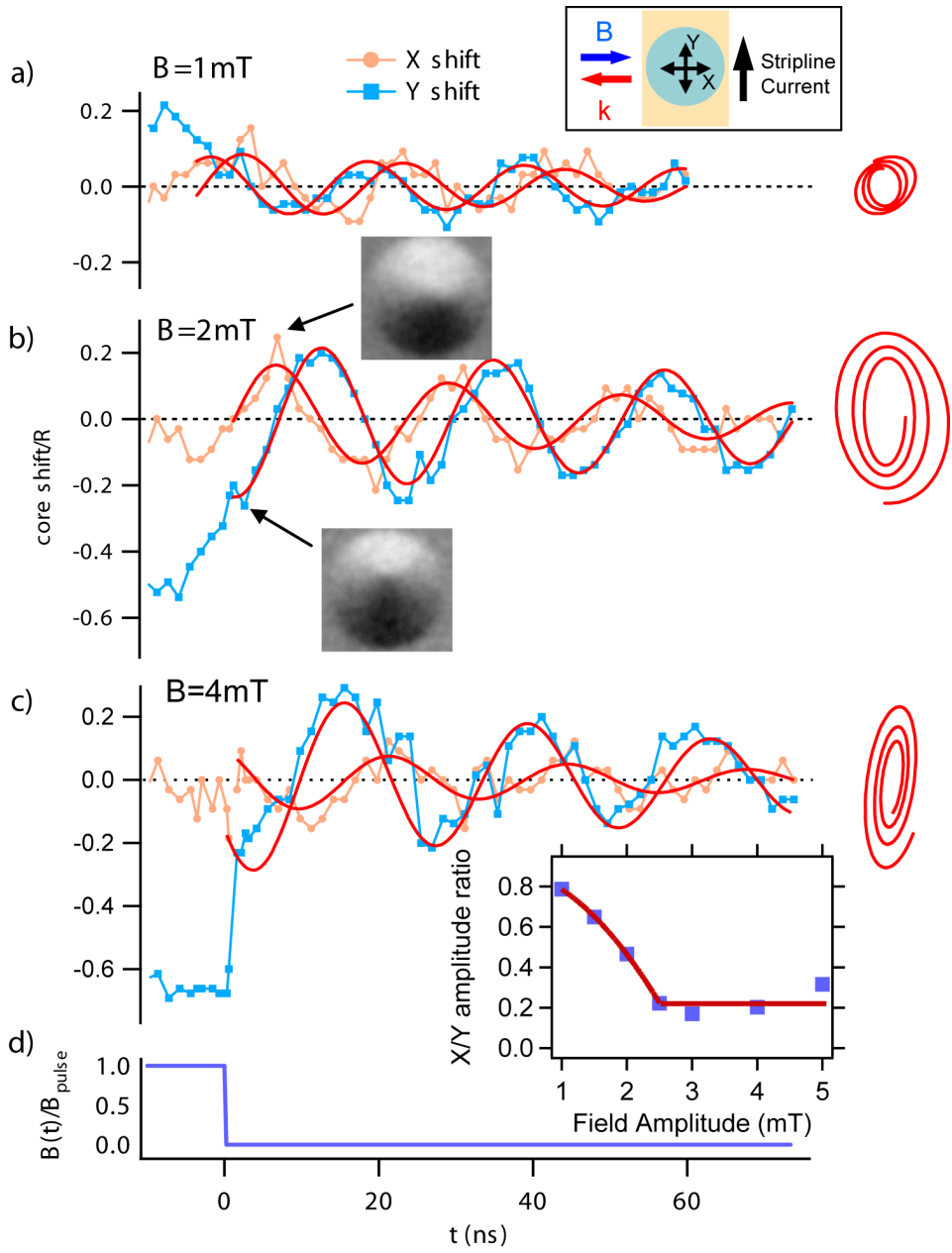

Figure 1.

Cheng et al.

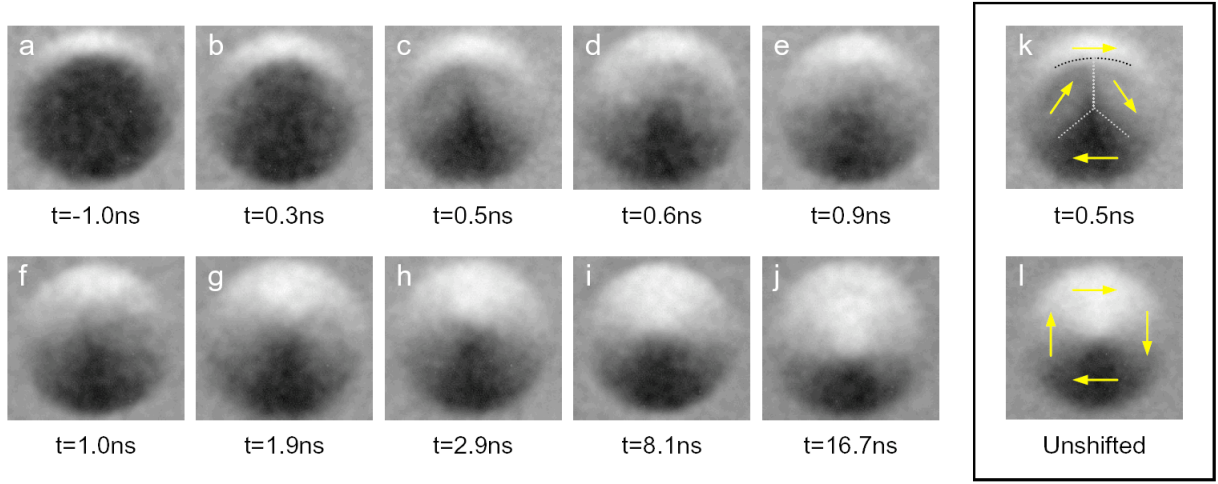

Figure 2.

Cheng et al.

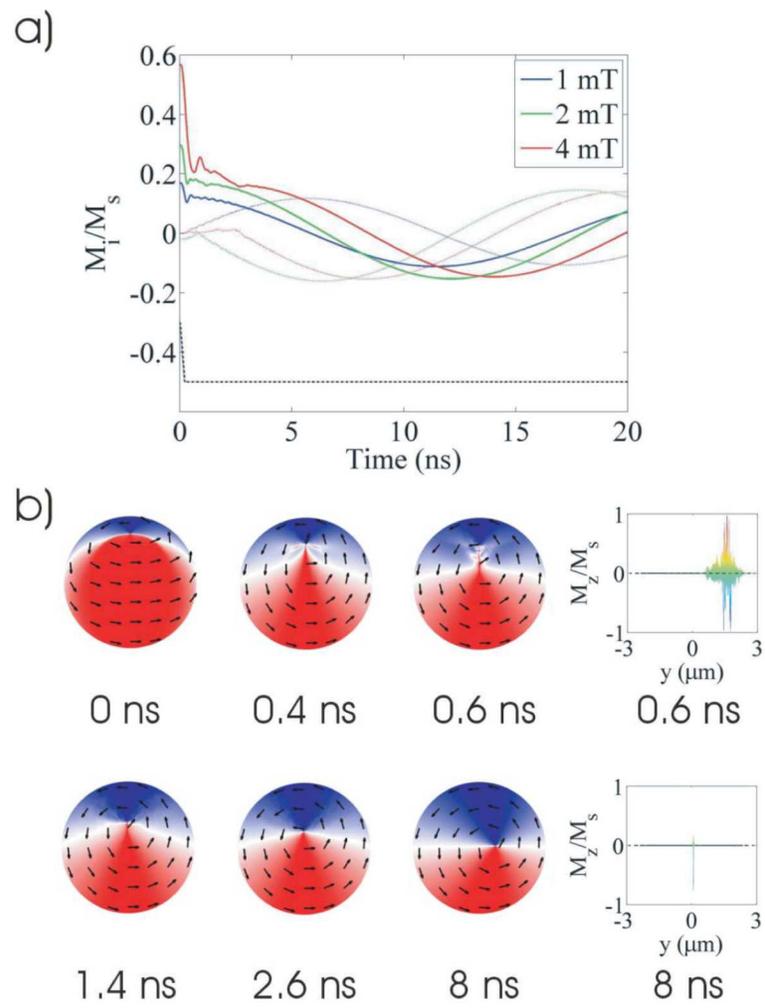

Figure 3.

Cheng et al.